\documentstyle[epsf]{article}
\newcommand{\vlargepspicture}[1]{\centerline{\setlength\epsfxsize{15cm}\epsfbox{#1}}}
\newcommand{\hpspicture}[1]{\centerline{\setlength\epsfxsize{17cm}\epsfbox{#1}}}

\newcommand{\err}[2]{\raisebox{0.08em}{\scriptsize{$\hspace{-0.8em}\begin{array}{@{}l@{}}
                     \plus\makebox[0.55em][r]{#1}\\[-0.15em]
                     \minus\makebox[0.55em][r]{#2}
                     \end{array}$}}}
\newcommand{\plus}{\makebox[15pt][c]{$+$}}
\newcommand{\minus}{\makebox[15pt][c]{$-$}}

\newcommand{\beq}{\begin{equation}}
\newcommand{\eeq}{\end{equation}}
\newcommand{\NP}{Nucl.~Phys. }

\title{\vspace{-3.0cm}
\hspace{10cm}{\normalsize HLRZ 96-20}\\
\vspace{-.3cm}
\hspace{10cm}{\normalsize WUB 96-17}\\
\vspace{1.5cm}
First Evidence of $N_f$-Dependence in the QCD Interquark Potential.}
\author{\underline{\sf SESAM-Collaboration}: \\ U.~Gl\"assner$^{\rm b}$, S.~G\"usken$^{\rm b}$, 
H.~Hoeber$^{\rm a}$, Th.~Lippert$^{\rm a}$, \\
        G.~Ritzenh\"ofer$^{\rm a}$,
        K.~Schilling$^{\rm a,b}$, G.~Siegert$^{\rm a}$, A.~Spitz$^{\rm
          b}$ and A.~Wachter$^{\rm a}$. 
\vspace{.3cm}\\
{\normalsize {\rm $^a$}HLRZ c/o KFA J\"ulich, D-52425 J\"ulich,
          and DESY, D-22603 Hamburg, Germany,}\\
{\normalsize{\rm $^b$}Physics Department, University of Wuppertal, D-42097
           Wuppertal, Germany.}
}       
\begin{document}
\maketitle
\begin{abstract}
We present a lattice calculation of the interquark potential between
static quarks in a ``full'' QCD simulation with 2 flavours of
dynamical Wilson-quarks at three intermediate sea-quark masses. We
work at $\beta = 5.6$ on lattice size of $16^3 \times 32$ with 
100 configurations per sea-quark mass. We compare 
the full QCD potential with its quenched counterpart at equal lattice
spacing, $a^{-1} \simeq 2.0$ GeV, which is at the onset of the quenched
scaling regime. We find that the full
QCD potential lies consistently below that of quenched QCD. 
We see no evidence for string-breaking effects on these lattice
volumes, $V \simeq (1.5\,\, {\rm fm})^3$. 
\end{abstract}
\section{Introduction\label{INTRO}}
Ever since the seminal paper of Creutz\cite{Creutz} on the confining
character of the static quark-antiquark interaction in the weak
coupling regime of quenched QCD, refined lattice methods have been
devised and targeted towards an improved determination of the
interquark potential. By now, quenched studies on large lattices
($\geq 32^4$) have reached a statistical accuracy in the percent
region at small and medium interquark separations, $R$, for 
bare lattice couplings in the scaling regime $6.0 \leq
\beta \leq 6.8$\cite{UKQCD,48,Potential}. This corresponds to lattice
resolutions in the range $1.94 \leq a^{-1} \leq 6$ GeV, which is fine
enough to observe running coupling effects\cite{amsterdam}\footnote{At
  present, such level of precision is extremely hard to realize in
lattice evaluations of hadronic attributes proper, even in quenched
QCD, since this requires very high statistics\cite{weingarten, ukawa,
gupta, schierholz}.}.  

As {\it full} QCD lattice simulations with Wilson fermions are presently
still in their infancy it is rather 
obvious to ask the question whether the form of the static potential
can provide a decent signal for fermion loop effects in the real QCD
vacuum state. The shape of the static potential is expected to grow
more convex as effects from dynamic fermions turn
gradually stronger with decreasing quark mass; 
however, this can only be observed provided we can 
access a large enough window of observation in
$R$ at sufficiently small quark masses.
We can
estimate the effect of dynamical fermions on the Coulomb-term of the
potential from the lowest-order coupling constant :
\beq
\alpha(R) \simeq -{1 \over 8 \pi}{1 \over b_0 \log{R \Lambda_V}}\,\, ,
\eeq
where the $N_f$-dependence is in $b_0$ ($N_f$ denotes the number of flavours):
\beq
b_0 = {33 - 2 N_f \over 48 \pi^2}\,\, .
\eeq
Naively, from $b_0$ only, we would expect to observe an effect on the order of $+$14
\% when switching from $N_f=0$ to $N_f=2$. Obviously, to be sensitive
to unquenching, one requires data with an overall error (statistical
{\it and} systematic) below the 5 \% level. Such a precision was
clearly not reached by previous lattice studies with dynamical Wilson 
fermions \cite{Gupta,hemcgc} in the scaling regime, which we 
expect to start at $\beta \stackrel{>}{\sim} 5.6. $\footnote{In this 
situation, one is tempted 
to assess the impact of dynamical fermions by matching, at
given values of $\beta$ and $\kappa_{sea}$, the Wilson loops
$W(R\times T)$ onto their respective quenched analogues, by suitable
$\beta$-shifts, $\Delta \beta (R\times T,\beta, \kappa_{sea})$. Using
this technique ref. \cite{Gupta} indeed found an increasing $\Delta
\beta$ with $R$. However, as this procedure can only be related to
physical quantities in the large $T$ limit one would prefer to
investigate the QCD potential directly \cite{Born}.} 
\par
In this paper, as part of our ongoing 100 Teraflopshrs simulation of
2-flavour Wilson-fermion QCD, which was described in Ref.\cite{Melb}, 
we will present a first high statistics study of the
interquark potential. We work at a bare lattice coupling of
$\beta=5.6$ on a lattice volume of $16^3 \times 32$ and simulate at
three sea-quark masses which correspond to the hopping parameter
values $\kappa_{\rm sea} = 0.156, 0.157$ and $0.1575$. With these
parameters the lattice resolution is expected to be around
$a^{-1}\simeq 2 $ GeV \cite{Gupta,Melb,Spectrum}.
\par
Field configurations are produced on Quadrics QH2 machines using the Hybrid
Monte Carlo Algorithm; for details we refer the reader to a
forthcoming publication\cite{Toolkit}. In
this analysis, we  exploit 2500 trajectories per
sea-quark mass\footnote{This is approximately half of our scheduled sample.}, 
which corresponds to three times 100 independent configurations (see section
\ref{AUTOCORRELATION}).  
\section{The Static Potential\label{POTENTIAL}}
\paragraph{Autocorrelation\label{AUTOCORRELATION}}
The statistical quality of a sample of configurations generated as an HMC
timeseries is largely determined by the integrated autocorrelation time
$\tau_{\rm int}$. 
We have obtained clean signals both for exponential and integrated
autocorrelation times of the plaquette using the complete sets of trajectories. In
table \ref{AUTOS} we list the values for the integrated
autocorrelation time $\tau_{\rm int}$.
\begin{table}[htb]
\begin{center}
\begin{tabular}{|cccc|}
\hline 
$\kappa_{\rm sea}$ & 0.156 & 0.1570 & 0.1575 \\ 
\hline
$\tau_{int}$ & 4.8(7) &  5.1(8) & 14.8(8) \\   
\hline
\end{tabular}
\caption{\label{AUTOS} 
Exponential and integrated autocorrelation times of the plaquette.}
\end{center}
\end{table}
Since the plaquettes are considered to represent the
worst case with respect to autocorrelations we choose to perform
measurements on configurations separated by 25 trajectories (a
detailed discussion of autocorrelation issues follows in a forthcoming
publication)\footnote{Note that
due to a variable step size ($N_{md}^{\rm mean} = 100$ and
$N_{md}^{\rm var} = 20$) these configurations are not equidistant in
molecular dynamics time.}.
\paragraph{Analysis Method}
The static (spin independent) potential $V(R)$ is computed in a
standard fashion 
from the path-ordered products of link variables around space-time
rectangles. In order to enhance the ground-state signal, we use a local
gauge-invariant APE-type \cite{APE} smearing procedure on the spatial
links. The smearing parameter $\alpha$ is set to $\alpha = 2$. We
have optimised the number $N$ of smearing iterations and find $N=20$
or $N=25$ to yield overlaps $\geq$ 80 \% for
all $\hat{R}$ with very little dependence on $\hat{R}$ (lattice
quantities are hatted to avoid confusion). 
\par
Similarly to the quenched case \cite{Potential}, off-axis measurements
are included for improved spatial resolution; this also allows an
investigation of rotational 
invariance restoration. We perform measurements at 36 different values
of $\hat{R}$ comprising 8 on-axis and 24 off-axis constellations.
\par
The potential is defined through local masses at time $\hat{T}$ : 
$\hat{V}^{\hat{T}}(\hat R) = \log{
 {{\cal W}(\hat R,\hat{T}) \over {\cal W}(\hat R,\hat{T}+1)}}$. 
We search for plateaus defined by
\beq
{\hat{V}^{\hat{T}-1}(\hat R) \over \hat{V}^{\hat{T}}(\hat R)} 
\simeq 1\,\, .
\eeq
Good plateaus are found for $\hat{T} \ge 3$. We have checked that 
local masses, $\hat{V}^{\hat{T}}(\hat R)$, and semi-local
masses, from constant fits of $\hat{V}^{\hat{T}}$ on three time-slices 
$\hat{T} = 3,4,5$, yield the same values
for the potential. Furthermore, the central values of the 
fit-parameters (as given below in 
tables \ref{fitparams} and \ref{STRING}) remain virtually unaffected:
however, using semi-local masses one may reduce the errors by
approximately 15 \% (albeit at the cost of a slightly increased $\chi^2$
per degree of freedom).
\par
Lattice artefacts in the potential due to the difference of the
one-gluon exchange on the lattice and in the continuum are corrected 
for by subtracting a term proportional to
\beq
\label{artefacts}
\delta \hat{V}(\hat R) = \left( [{1 \over \hat R}] - { 1 \over \hat R}
  \right)\,\, ,
\eeq
where $[{1 \over \hat R}]$ denotes the lattice one gluon exchange
extrapolated to infinite volume \cite{Potential,Michael,Rebbi}:
\beq
[{1 \over \hat R}] = 4 \pi \, {\rm lim}_{V \rightarrow \infty} G(\hat
R; V)\,\,\, \mbox{   with   }\,\,\, G(\hat R; V) = { 1 \over V} \sum_{\hat{k}_i} {
  \prod_i {\sf cos} (\hat{k}_i \hat R_i)
  \over 4 \, \sum_i {\sf sin}^2 (\hat{k}_i/2) }\,\, ,
\eeq
where $\hat{k}_i = {2 \pi \over L}m_i$, $m_i = -L/2 + 1, \ldots, L/2$,
  $V=L^3$. 
\par
In order to fix the size of the correction term $g\, \delta \hat{V}$ we interpolate our data 
using the following four-parameter ansatz (the detailed form of the fit
is of no interest here, we only wish to show that a smooth
interpolation can be obtained):
\beq
\label{standard}
\hat{V}(\hat{R})= \hat{V}_0+ \hat{k}\hat{R}- {e \over \hat{R}}+g \, \delta \hat{V}(\hat{R})\,\, .
\eeq
\paragraph{Results}
\begin{figure}[h]
\vspace{-8cm}
\vlargepspicture{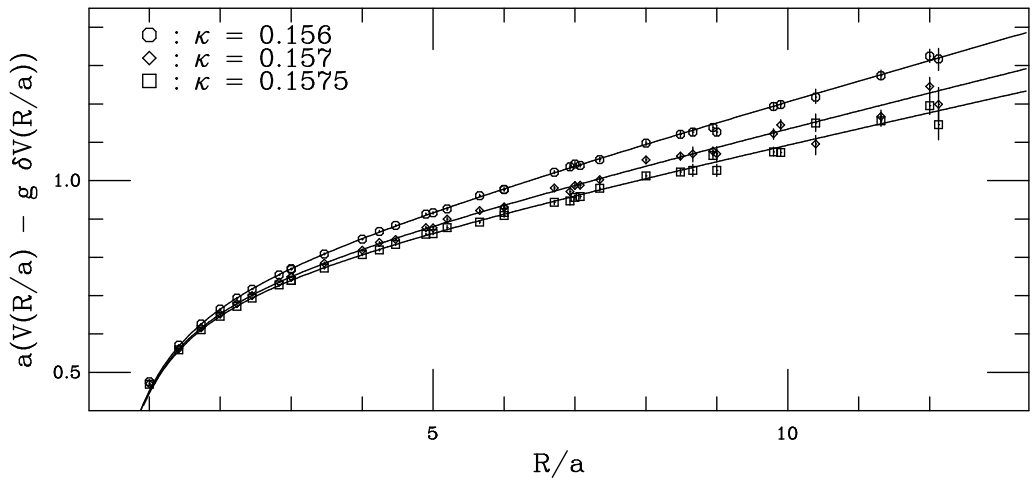}
\caption{\label{dataplot}
$\left( \hat{V}(\hat R) - g \delta
  \hat{V}(\hat R) \right)$ as a function of the lattice separation $\hat
  R$.}
\end{figure}
The data and fits are shown in
figure \ref{dataplot} where we plot $\left( \hat{V}(\hat R) - g \delta
  \hat{V}(\hat R) \right)$ as a function of the lattice separation $\hat
R$. Throughout the analysis we exclude the two smallest $\hat R$
values from the fit\footnote{These points are, however, included in
  figure \ref{dataplot}.}. The resulting parameters are given in table
\ref{fitparams}; errors are obtained from a bootstrap sample of size
250 and correspond to 68 \% confidence level. 
We find all $\chi^2$ per degree of freedom to be around 1. 
\begin{table}[hbt]
\begin{center}
\begin{tabular}{|c|ccccc|}
\hline
$\kappa_{\rm sea}$ & 
$\hat{V}_0$ & $\hat{k}$ & $e$ & $g$ & $\chi^2 /$ d.o.f \\ \hline 
0.156  &  0.724 \err{7}{6} & 0.0514 \err{13}{12} & 0.324 \err{9}{8} &
0.36 \err{1}{2} 
& 18/30 \\ 
0.157  &  0.721 \err{8}{6} & 0.0445 \err{10}{11} & 0.316 \err{11}{8}  &
0.35 \err{2}{2} 
& 31/30 \\ 
0.1575 &  0.728 \err{7}{7} & 0.0396 \err{13}{11} & 0.323 \err{10}{10}  &
0.37 \err{2}{2} 
& 25/30 \\ \hline 
quenched &   & & & & \\ \hline
6.0 &  0.6591 \err{43}{43} & 0.04771 \err{81}{81} & 0.2917 \err{59}{59}  &
0.2916 \err{74}{74}& 10/29 \\ \hline
6.2 &  0.6336 \err{24}{24} & 0.02574 \err{34}{34} & 0.2865 \err{46}{46}  &
0.265 \err{11}{11}& 43/63 \\ \hline
\end{tabular}
\caption{\label{fitparams}Fit results from the ansatz
  eq. \ref{standard}. For a discussion of the quenched data see
  section \ref{discussion}.
}
\end{center}
\end{table}
\par
\begin{figure}[tb]
\unitlength 1.05cm
\begin{center}
\vspace{-1cm}
\begin{picture}(14,7)
\put(0,0){\epsfxsize=7cm\epsfbox{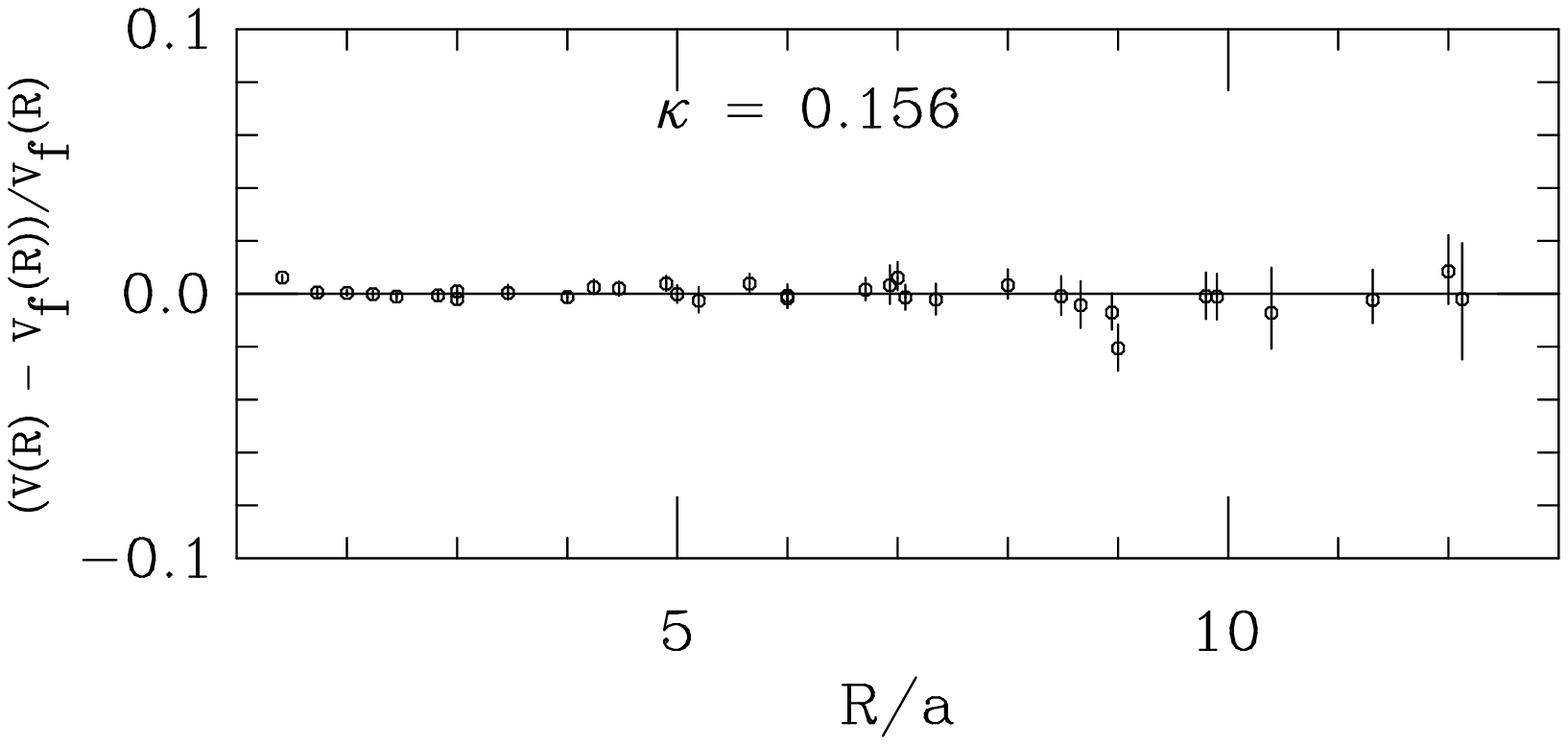}}
\put(7,0){\epsfxsize=7cm\epsfbox{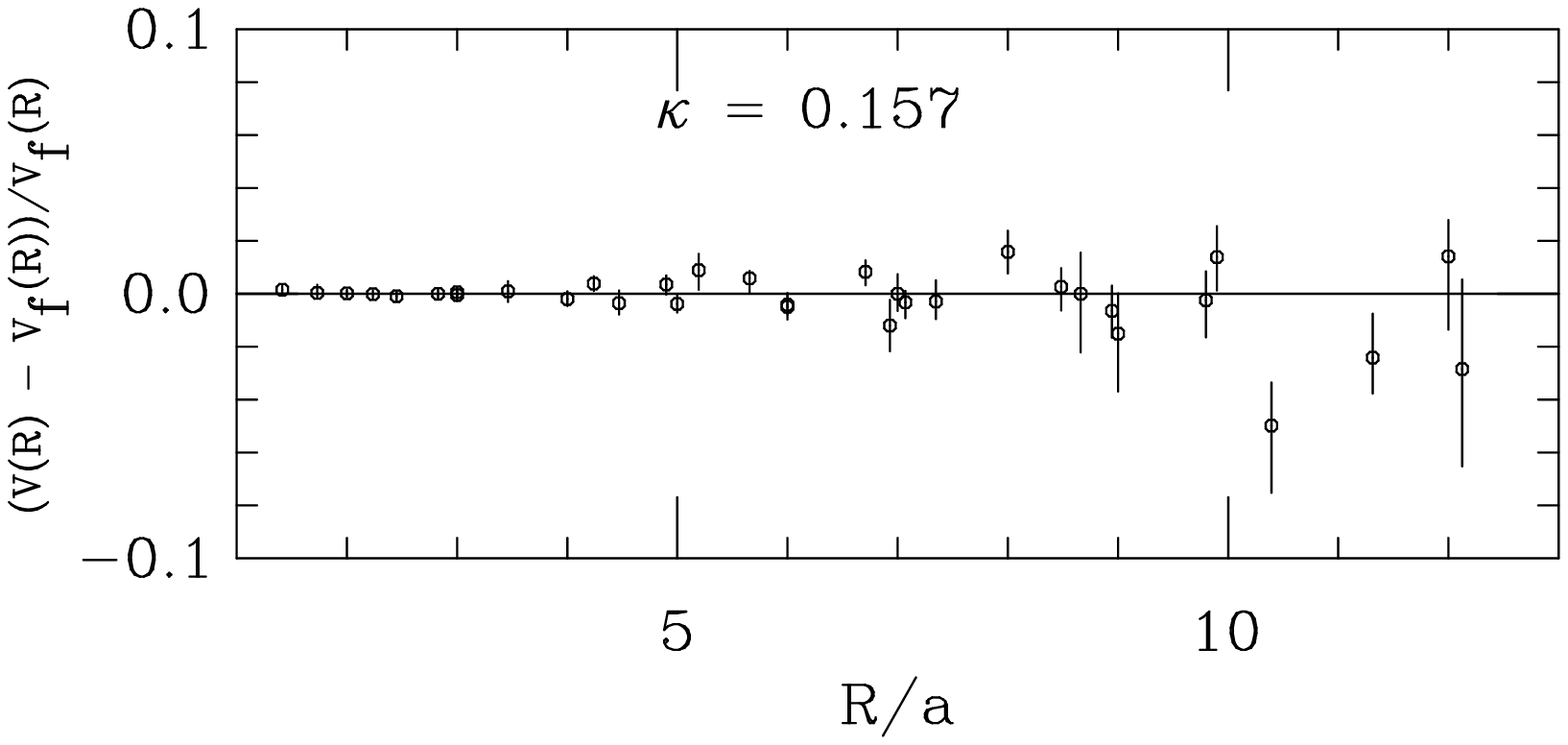}}
\put(3.0,3.5){\epsfxsize=7cm\epsfbox{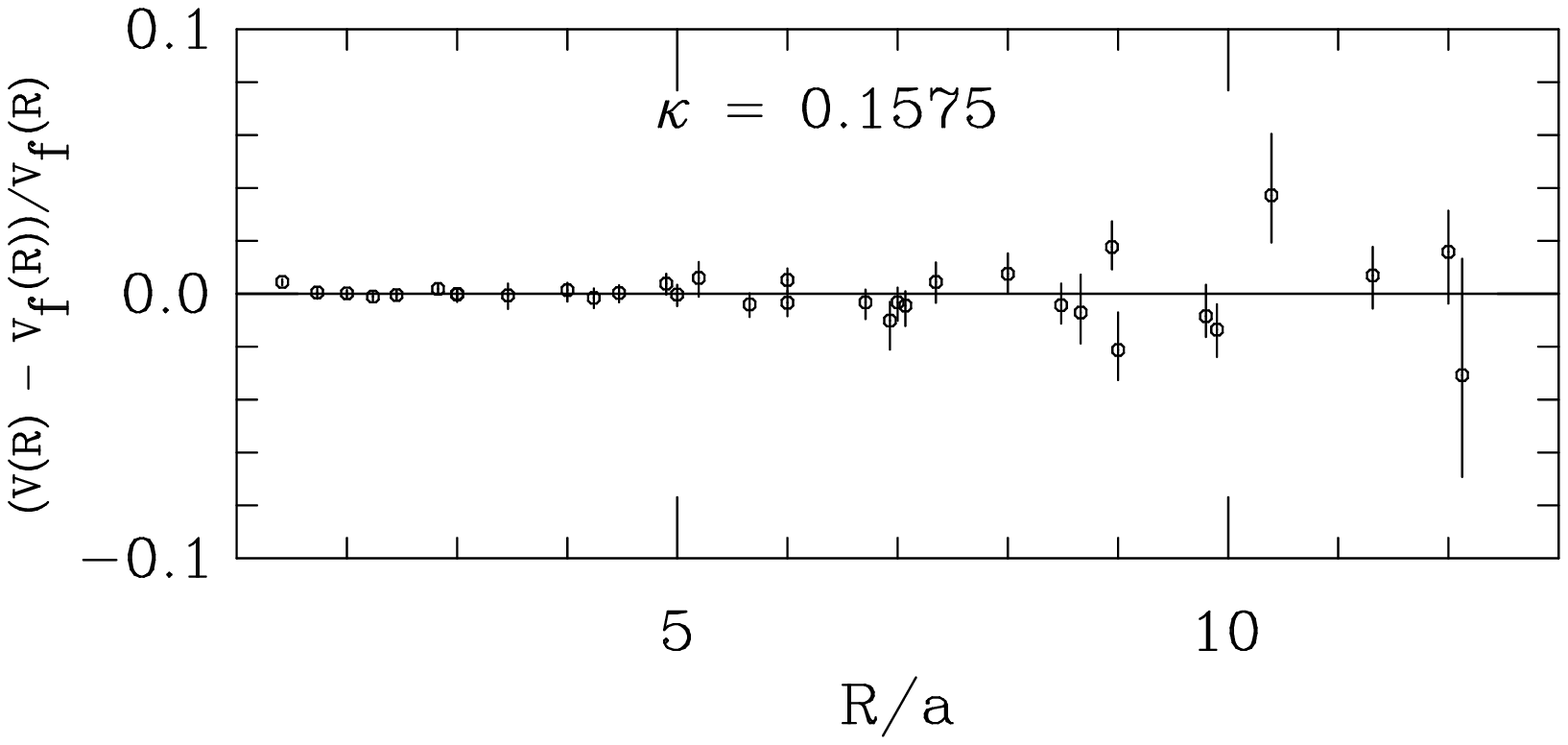}}
\end{picture}
\end{center}
\vspace{-1cm}
\caption{Relative deviations $(\hat{V}(\hat R)-V_f(\hat R))/V_f(\hat R)$ 
of the potential values from the fit curves $V_f(\hat R)$, eq. \ref{standard}.}
\label{errorplot}
\end{figure}

A more detailed view of the quality of the data and the
parametrisation can be obtained by
looking at the relative deviations of the data points from the
interpolating curves as shown in figure \ref{errorplot}. We find
the deviations to be smaller than $\pm 2$\% with minor exceptions at
very large $\hat{R}$. 
We conclude that adding a term proportional
to $\delta \hat{V}(\hat R)$ is sufficient to remove lattice artefacts at
small $R$ and that our interquark potentials
show no sign of non-linear behaviour towards large $R$, as predicted
from string breaking. 
\section{Scaling the data\label{scaling}}
In the case of full QCD it is appropriate to find the lattice scale
from a quantity defined at an intermediate distance rather than using
the asymptotic quantity ``string-tension''. Therefore, 
following ref. \cite{Sommer}, we extract the lattice spacing $a_{\rm
  F}$ by matching the force from our lattice potential to
the force obtained from potential models as applied to quarkonium
  spectroscopy. This amounts to finding the numerical solution, in
  terms of $\hat{R}_0$, of the equation 
\beq
\hat{R}_0^2 {d\hat{V} \over d\hat{R}}_{\mid \hat{R}_0} = 1.65\,\, ,
\eeq
which, in phenomenological models, is given by the reference length
\beq
R_0 = a_{\rm F} \hat{R}_0 = 0.5\,\, {\rm fm}\,\, .
\eeq
Since this procedure involves the use of a numerical derivative it is clearly
  more delicate than setting the scale by the string-tension and hence
  requires
  precise data. 
\par
The force values are calculated from the finite differences
\beq
\label{barR}
\hat{F}(\bar{R}) = {\hat{V}_c(\hat{R}_2) - \hat{V}_c(\hat{R}_1) \over (\hat{R}_1 -
\hat{R}_2)}\,\, ,
\eeq
where $\hat{V}_c(\hat{R})$ are the corrected data which we view as our
lattice estimates of the continuum QCD potential :
\beq
\hat{V}_c(\hat{R}) = \hat{V}(\hat{R}) - g \delta \hat{V}(\hat{R})\,\, .
\eeq
All possible pairs of potential data with distances $1.4 \geq 
(\hat{R}_1-\hat{R}_2) \geq 1$ are used to calculate an improved
$\bar{R}$ in eq. \ref{barR} by demanding that 
\beq
\label{rbar}
-{\partial V_{{\rm theo}}(\bar{R}) \over \partial \bar{R}} 
= F(\bar{R})\,\, ,
\eeq
where 
\beq
V_{\rm theo}(R) = V_0 + k\,R - {e \over R}\,\, .
\eeq
Eq. \ref {rbar} is then solved for $\bar{R}$ using the fit parameters 
of table \ref{fitparams}. 
\par
The resulting values $\hat{R}_0$ and lattice
spacings $a_{\rm F}$ are given in table \ref{STRING}. 
We find very little difference between the values of $\hat{R}_0$
obtained from this numerical difference procedure and those obtained
analytically, with the fit parameters of eq. \ref{standard} :
\beq
\hat{R}_0 = \sqrt{{1.65 - e \over \hat{k}}}\,\,  .
\eeq
This is taken as further evidence of the smoothness of our data and as an
indication for the stability of the method.
\begin{table}[hbt]
\begin{center}
\begin{tabular}{|c|ccc|}
\hline
$\kappa_{\rm sea}$ & 
$\hat{R_0}$ & $a^{-1}_{\rm F}[GeV]$ & 
$a^{-1}_{\rm V}[GeV]$ \\ \hline 
0.156  &  5.13 \err{7}{8} & 2.02 \err{3}{3} &  1.94 \err{2}{2}  \\ 
0.157  &  5.55 \err{17}{17} & 2.19 \err{7}{7}  &
2.09 \err{3}{2}  \\ 
0.1575 & 5.83 \err{11}{12}  & 2.30 \err{4}{5} &
2.21 \err{3}{3}  \\ \hline 
quenched & & &  \\ \hline 
6.0 & 5.35 \err{2}{3}  & 2.11 \err{1}{1} &
2.02 \err{1}{1}  \\ \hline 
6.2 & 7.27 \err{3}{3}  & 2.86 \err{1}{1} &
2.74 \err{1}{1}  \\ \hline 
\end{tabular}
\caption{\label{STRING}Lattice spacings obtained from
the force at $R_0 = 0.5$ fm and from the string tension $\hat{k}
= a^2 \times 0.1936 \,{\rm GeV}^2$ (see table \ref{fitparams} for $\hat{k}$). Errors
are statistical only. The quenched data are discussed in section \ref{discussion}.}
\end{center}
\end{table}
Note that the statistical errors are as low as 3 \%.
\par
As we see no deviation of the potential from a linear behaviour for
large $R$ we can also attempt to extract the lattice spacing from the force
at larger distances, i.e. at around 1 fm, where our data are dominated
by the linear term in eq. \ref{standard}. In practical terms, we then 
interpret the values of $\hat k$ from table
\ref{fitparams} as string-tension values from which we derive the
lattice spacings $a_{\rm V} = \sqrt{\hat k / \sigma}$, where we
take, somewhat arbitrarily, $\sqrt{\sigma} = 0.44\, {\rm GeV}$ (see
table \ref{STRING}). Note that the sizes of the statistical errors are beginning to
be competitive to those of previous quenched calculations (without link-integration)
\cite{Potential,Wittig}.  
Moreover, we observe that the ratio $a_{\rm F}/a_{\rm V}$ remains
unchanged for all three sea-quark masses indicating that a slightly different
choice of $\sqrt{\sigma}$, which we find to be 458 $\pm 12$ MeV for
$\kappa = 0.1575$, renders
consistent lattice spacings from both schemes.
\par
We comment that our hadron spectrum analysis yields lattice spacings
consistent with those of table \ref{STRING}, when using the mass of
the $\rho$ to set the scale \cite{Spectrum}.
\par
\begin{figure}[h]
\vspace{-8cm}
\hpspicture{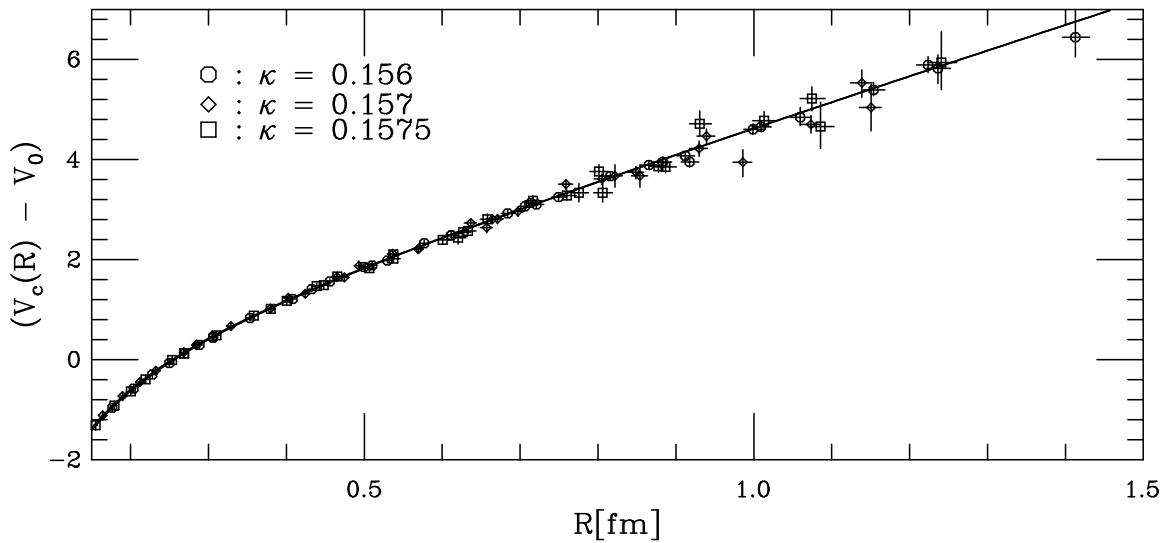}
\caption{\label{spots}
Scaled potential (using $a_{\rm V}$) in physical units. 
}
\end{figure}

In figure \ref{spots} we show our results in scaled form. We 
find that the data collapse to a universal
potential over the complete range of our measurements, $R \in
[0.1,1.5]$ fm. 
An analysis similar to that of figure \ref{errorplot} shows that nearly
all points lie within $\pm 2$ \% of the universal curve.
\section{Discussion : effects of dynamical fermions ? \label{discussion}}
At first sight, it might seem disappointing to find that our scaled
data collapse to a universal function; clearly, this shows that the
variation of $\kappa_{\rm sea}$ from 0.156 to 0.1575 is too small to
resolve mass-dependence in the static potential. However, we can do more/better
by comparing our data to the extreme case, the case of infinitely heavy sea
quarks, that is, the quenched approximation. 
\par
Table \ref{STRING}
shows that the lattice spacings in the present calculation correspond
to quenched couplings slightly above $\beta = 6.0$. We 
therefore performed two high statistics runs in
pure gauge theory including link-integration at $\beta = 6.0$ (lattice
size $16^4$, 570 configurations)
 and $\beta = 6.2$ (lattice size $32^4$, 100 configurations)
 \cite{Wachter}, the $\beta=6.2$ simulation being done to investigate scaling.
In order to achieve comparable systematic errors the evaluation of
the quenched potential data and their analysis follow the same procedure as
described in section 2 and we perform four-parameter fits at both $\beta$-values 
using a common lower cut $R > 0.196$ fm, designed to give
good $\chi^2/d.o.f.$ for all fits\footnote{We have checked however,
  that varying the fit
  range to include more values of $V(R)$ at smaller $R$ leaves our
conclusions largely unaffected. The fit parameters for the full QCD case
  are, in fact, completely stable.}. 
\par
Results are shown in tables \ref{fitparams} and \ref{STRING}. 
We find the quenched data, whose lattice spacings vary by about 40 \%,
to fall upon a universal curve with
deviations smaller than 1 \%, corroborating the earlier results of
ref \cite{Potential}. Furthermore, since the lattice spacings at
$\beta_{N_{f}=0}=6.0$ roughly correspond to those of $\beta_{N_{f}=2}=5.6$ (see
table \ref{STRING}) we are indeed at the onset of the scaling regime
for our full QCD simulation.
\par
\begin{figure}[h]
\vspace{-3cm}
\vlargepspicture{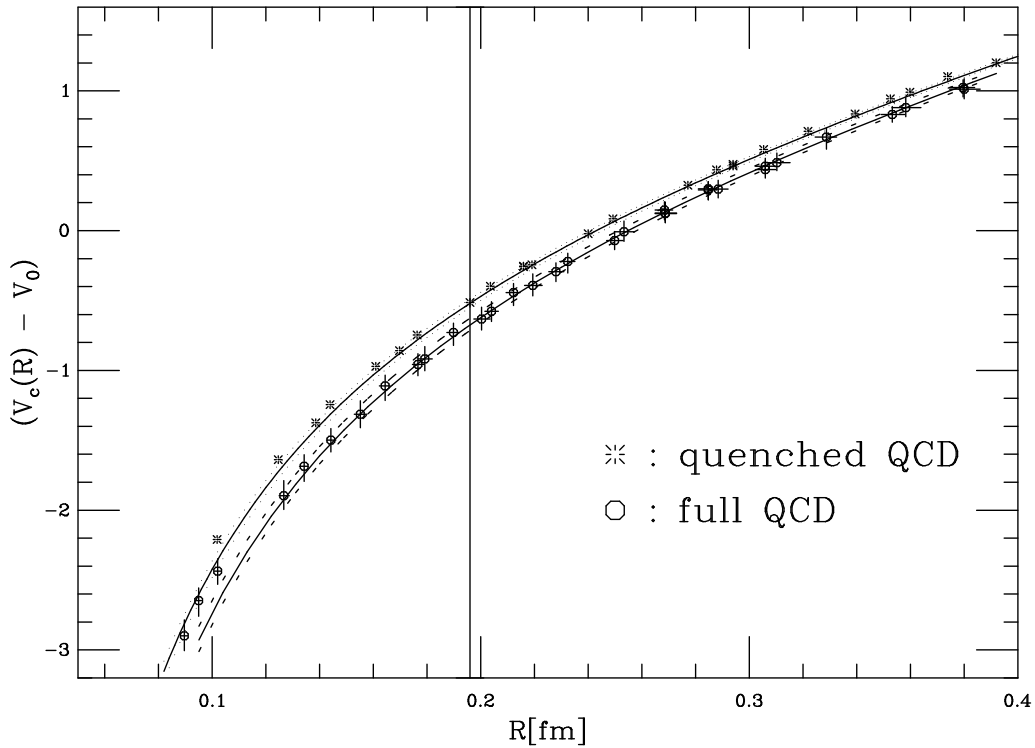}
\caption{\label{running}
Comparison of scaled potentials from quenched and full QCD (combined
data for quenched ($\beta=6.2$ and $\beta=6.0$) and full (three
sea-quark values) QCD). The plot shows both fits with their one-sigma
errorbands where only points right of the vertical line were used for
fits, $R \in [0.196, \infty]$ fm.}
\end{figure}

The comparison of quenched and unquenched 
QCD is shown in figure \ref{running} where we
plot the data in the small $R$-region, $R < 0.4$ fm. 
We find the full QCD data to lie significantly below that of quenched 
QCD ! In this scaling representation the data are solely
sensitive to the strength of the Coulombic term $e$. In order to
quantify the effect of dynamical fermions we can therefore directly
extract the information from the $e$-values as quoted in table
\ref{fitparams}. As for both cases, full and quenched QCD, we observe
good scaling, we can represent each data-set by its appropriate
average value, $e_{N_{f}=2} = 0.3210(100)$ and
$e_{N_{f}=0}=0.2890(55)$. 
From these numbers, we find unquenching to induce a decrease of 11 \%
in the coulombic part of the static potential.
\section{Summary and conclusions}
We have presented a calculation
of the static potential which is precise enough to unravel, for the
first time, the effects of unquenching at intermediate
distances.  We find the full QCD coupling constant to lie consistenly
above that of quenched QCD. However, the data are not sensitive enough
to see variation of the potential with the sea-quark mass. Our method 
is based on a consistent modelling of the $N_f=2$ and $N_f=0$
potentials. The full QCD lattice spacings are in the range $a^{-1}
\simeq 2$ GeV which, in the quenched case, is at the onset of the
scaling regime, $\beta_{N_{f}=0}\simeq 6.0$. 
Our result is in accord with the $N_f$-dependence of $\alpha$
as extracted from the perturbative expansion of the plaquette (see for
example \cite{Aoki}).
\par
The lattice volume of $(1.5\,\, {\rm fm})^3$ was not large enough to observe string
breaking. Note that using quenched data, one can obtain an 
upper bound for the distance at which string breaking should
be observable to be at $1.7$ fm \cite{Static}.
\par
We are currently performing a simulation with the same
lattice-resolution but 50 \% extended lattice size, in order to assess
finite-size effects and to work at lighter quark masses \cite{chiral}.
\par
We are encouraged by these results in our systematic search for Sea quark
Effects on Spectrum And Matrix elements.
\paragraph{Acknowledgements}
We are grateful to DESY and to the DFG for granting substantial
amounts of computer time on their QH2 Quadrics systems (at DESY/IfH 
in Zeuthen and the University of Bielefeld) which was instrumental to creating 
the QCD vacuum field configurations, and to the
IAA of the Bergische Universit\"at Wuppertal, where we could use the CM5
to extract the QCD potential data. Our project and the CM5 in Wuppertal
have both been supported by Deutsche Forschungsgemeinschaft through grants
Schi 257/1-4 and 257-3-2(3).  The quenched computations have partly
been done on the 64 node CM5 system of the GMD.  TL and KS appreciate
support by EU contract CHRX-CT92-0051 which enabled us to build up a
cooperation with our colleagues from the INFN Pisa and Roma whom we
admire for building, with APE, a very functional hardware for doing
computational physics.

\end{document}